%% file: paper.tex
\PassOptionsToPackage{table}{xcolor}
\documentclass[sigconf]{acmart}
\input{packages}

\setcopyright{rightsretained}
\copyrightyear{2024}
\acmYear{2024}
\acmDOI{XXXXXXX.XXXXXXX}

\acmConference[Conference acronym 'XX]{Make sure to enter the correct
  conference title from your rights confirmation emai}{June 03--05,
  2018}{Woodstock, NY}
\acmISBN{978-1-4503-XXXX-X/18/06}

\begin{document}

\title{Security Advice for Parents and Children About Content Filtering and Circumvention as Found on YouTube and TikTok}
	
\author{Ran Elgedawy}
\email{relgedaw@vols.utk.edu}
\affiliation{%
	\institution{University of Tennessee, Knoxville}
	\country{USA}
}

\author{John Sadik}
\email{johns@utk.edu}
\affiliation{%
	\institution{University of Tennessee, Knoxville}
	\country{USA}
}

\author{Anuj Gautam}
\email{agautam1@vols.utk.edu}
\affiliation{%
	\institution{University of Tennessee, Knoxville}
	\country{USA}
}

\author{Trinity Bissahoyo}
\email{tbissaho@vols.utk.edu}
\affiliation{%
	\institution{University of Tennessee, Knoxville}
	\country{USA}
}

\author{Christopher Childress}
\email{cchildr3@vols.utk.edu}
\affiliation{%
	\institution{University of Tennessee, Knoxville}
	\country{USA}
}

\author{Jacob Leonard}
\email{jleona19@vols.utk.edu}
\affiliation{%
	\institution{University of Tennessee, Knoxville}
	\country{USA}
}

\author{Clay Shubert}
\email{cshubert@vols.utk.edu}
\affiliation{%
	\institution{University of Tennessee, Knoxville}
	\country{USA}
}

\author{Scott Ruoti}
\email{ruoti@utk.edu}
\affiliation{%
	\institution{University of Tennessee, Knoxville}
	\country{USA}
}

\begin{abstract}
  \input{abstract}
\end{abstract}

\ccsdesc[500]{Human-centered computing~Empirical studies in HCI}
\ccsdesc[500]{Security and privacy~Social aspects of security and privacy}

\keywords{Security advice, parents, children, privacy, content filtering}

\maketitle

\input{intro}
\input{rw}
\input{method}
\input{result}
\input{goals}

\input{discussion}
\input{conclusion}

\bibliographystyle{ACM-Reference-Format}
\bibliography{latex/bibtex/publications,paper}

\appendix
\input{appx_codebook}

\end{document}

%% file: packages.tex
\makeatletter\def\input@path{{latex/styles/}}\makeatother 
\usepackage{fast-todo}
\usepackage{formatting}

\usepackage{adjustbox}
\usepackage{multirow}
\usepackage{booktabs}
\usepackage{tabularx}

%% file: abstract.tex
In today's digital age, concerns about online security and privacy have become paramount.
However, addressing these issues can be difficult, especially within the context of family relationships, wherein parents and children may have conflicting interests.
In this environment, parents and children may turn to online security advice to determine how to proceed.
In this paper, we examine the advice available to parents and children regarding content filtering and circumvention as found on YouTube and TikTok.
In an analysis of 839 videos returned from queries on these topics, we found that half (n=399) provide relevant advice.
Our results show that of these videos, roughly three-quarters are accurate, with the remaining one-fourth containing factually incorrect advice.
We find that videos targeting children are both more likely to be incorrect and actionable than videos targeting parents, leaving children at increased risk of taking harmful action.
Moreover, we find that while advice videos targeting parents will occasionally discuss the ethics of content filtering and device monitoring (including recommendations to respect children's autonomy) no such discussion of the ethics or risks of circumventing content filtering is given to children, leaving them unaware of any risks that may be involved with doing so.
Ultimately, our research indicates that video-based social media sites are already effective sources of security advice propagation and that the public would benefit from security researchers and practitioners engaging more with these platforms, both for the creation of content and of tools designed to help with more effective filtering.

%% file: intro.tex
\section{Introduction}\label{intro}

In today's digital age, concerns about online security and privacy have become paramount.
However, addressing these issues can be difficult, especially within the context of family relationships, such as those between parents and children.
Parents face the daunting task of safeguarding their children online while still respecting their children's rights.
Children, in turn, seek ways to ensure their autonomy and in extreme cases escape from abusive home environments.
This delicate balance between each party's interests can make it difficult to identify an appropriate path forward.

To help navigate this situation, it is understandable that parents and children would turn to online information sources to better understand these issues and the technological resources available to them for achieving their respective goals.
However, there is little research that evaluates the types of online security advice available to parents and children regarding online content filtering.
Key questions regarding this advice include, 
(i) what advice is being provided to parents and children,
(ii) is the information presented accurate and actionable,
and (iii) is there a balanced discussion of the competing interests of each party?

To shed light on these questions, we examine security advice found on the video-based social media platforms of YouTube and TikTok.
We focus on these sources as prior research has shown that they are becoming common places to share advice~\cite{medinaserrano2020dancing,wei2022anti,basch2022covid}.
On each platform, we executed 33 search queries on the topics of content filtering and circumvention.
This resulted in 839 videos, each of which we viewed and analyzed.
Of those videos, slightly less than half (n=399) turned out to be relevant to the topics of content filtering and circumvention.
For relevant videos, we analyzed them based on their target audience, topical content, accuracy, actionability, and how they discuss the interplay between parental and child rights.

Key findings from our video analysis include, 

\begin{itemize}
	\item
	We find that roughly three-quarter of security advice videos contain correct, comprehensive, and actionable content. 
	This indicates that these platforms can be valuable sources of security advice for parents and children.
	However, with nearly one-quarter of videos containing inaccurate content, there is a need for more research into how to assist either the platforms or parents/children to effectively filter the videos they are presented with.
	
	\item
	Our analysis reveals an imbalance in the quality of videos targeting parents and children.
	For parents, videos are highly accurate (91\%), but less likely to be actionable (71\%)
	In contrast, videos targeting children are highly actionable (92\%), but less likely to be accurate (77\%).
	While the former is less than desirable, the later is more dangerous, as the combination of actionable but inaccurate advice could have negative ramifications for the children who follow it.
	
	\item 
	Our analysis discovers that one in ten videos made for parents discuss the ethics of content filtering and device monitoring, including a discussion of why such protections may be inappropriate.
	This is an encouraging result as it means that parents are more likely to think through the implications of implementing content filtering and device monitoring technologies.
	In contrast, no videos attempted to teach children why parents may be trying to filter content or the risks of circumventing protections.
	This situation has the potential to leave children at increased risk for unintended consequences as they circumvent parental protections.
\end{itemize}

Ultimately, our research indicates that video-based social media sites are already effective sources of security advice propagation, particularly for children and to a lesser extent parents.
However, based on our results, we also believe that the public would benefit from security researchers and practitioners engaging more with these platforms.
First, since security advice videos are being engaged with on these platforms, we think it would be wise for experts to create and publish content on these platforms, increasing the quality of security advice found therein.
Second, we think there is room for more effective tooling to help users identify factually relevant and correct videos.
This could take the form of researching how to build better search filtering tools or identifying mechanisms for the community to more easily label misleading content.

%% file: rw.tex
\section{Related Work}\label{rw}

\subsection{Tiktok as a Source of Qualitative Data}

While YouTube has long been a platform for sharing information and experiences, TikTok represents a newer and rapidly growing medium for creative expression.
With over 1 billion monthly users~\cite{tiktok}, TikTok's influence and reach are undeniable, particularly among parents and children.
As of early 2022, 35\% of TikTok's users are between 18 and 24 years old and an additional 14\% are under 18~\cite{stats}, which is very relevant to the targeted age group in our work.

While TikTok is relatively new, prior research has already used it as an information source.
In the area of security and privacy, De Leyn et al.~\cite{deleyn2022childs} investigated how tweens (kids between 8-12 years old)  and their parents perceive and manage risks and opportunities on TikTok, including privacy risks.
Wei et al.~\cite{wei2022anti} explored the types of advice given on TikTok related to device monitoring for intimate partner and child-parent relationships.
TikTok has also been used as an information source for exploring information sharing about COVID-19~\cite{basch2022covid} and politics~\cite{medinaserrano2020dancing}.

\subsection{Interpersonal Security and Privacy}

There is already a significant body of work examining parent-child interactions and perceptions within the context of security and privacy.
For example, research has examined the types of information parents share online and how this might reveal information that children don't want shared~\cite{blumross2016sharenting,brosch2018sharenting}.
Similarly, there is work investigating the attitudes of parents and teens towards monitoring children and their devices~\cite{leaver2017intimate,steeves2010editorial,cranor2014parents,czeskis2010parenting}.
Unsurprisingly, this work shows that parents are more likely than not to see value in monitoring, while teens are generally averse to monitoring.
There have also been efforts to explore how IoT devices (e.g., smart locks and speakers) can lead to conflicts between parents' desire for control to ensure the safety of their kids and the kids' desire for privacy \cite{he2018rethinking,lau2018alexa}.

Looking at intimate partner relationships, Chatterjee et al. \cite{chatterjee2018spyware} investigated spyware apps used for intimate partner surveillance.
Likewise, Roundy et al. \cite{roundy2020many} investigated creepware, apps whose primary use is allowing non-technical users to launch interpersonal attacks.

Largely missing from the above research is an examination of information sources related to interpersonal security and privacy.
To start filling this knowledge gap, Wei et al.~\cite{wei2022anti} examined TikTok videos discussing advice for setting up monitoring apps, showing that there were a fair number of such videos.
While Wei et al.'s work focused on intimate partner monitoring, they also found 26 videos on the topic of parent-child monitoring.
\emph{In our research, we expand upon the work of Wei et al. by focusing on the question of parent-child security advice.}

\subsection{Content Filtering}
Content filtering in the context of a parent-child relationship has been studied before. 
Hashish et al. worked on a collaborative model where parents and children worked together to set constraints and filters on the children's devices \cite{hashish2014involving}. 
This allowed the parents to talk to their children about content filtering. 
Specifically, the app provided an entry point to the discussion about content filtering, but it also helped parents understand how much their kids understood about what is appropriate. 
It also helped parents gauge their kids' interests. 
While this app allowed collaboration in setting the guidelines, parents tended to still desire the final say on the content their kids interacted with, even if that content was identified as content intended for children of the appropriate age. 

This collaborative model is important because the work of Ghosh et al. identified that strict parents, those who gave their teens less freedom, were most likely to use parental control apps for their teens \cite{ghosh2018matter}.
Ghosh et al. presented the first study to look at the characteristics of the parents and teens (ages 13 to 17) who use parental control apps. 
In addition to the above conclusion, they also found that teens who were victimized online or had peer problems were more likely to be monitored by their parents. 
However, importantly, they found that increased parental control for teens was associated with more online risks. 
In fact, they found little evidence to support the idea that using parental control apps protected teens from experiencing online risks. 
Ghosh et al. suggest that this could be due to the parental control apps reinforcing bad parenting habits that were harmful to their children. 
Finally, they found that parents often overestimated how much they involved their teens in setting boundaries, and parents also overestimated how autonomy-granting they were. 
Meanwhile, teens felt like their parents were not involving them very much and were not very autonomy-granting. 
This is consistent with prior research from Blackwell et al. \cite{blackwell2016managing}, as Ghosh et al. mention. 

Similarly, the work of Wang et al. noted that the overuse of restrictions and surveillance, like through an app on the child's phone, can put a strain on the parent-child relationship, leading to unintended negative consequences \cite{wang2021protection}.
Wang et al. analyzed the different features of parental control apps that have been implemented. 
They focused their analysis on three axes: granularity, feedback/transparency, and parent-child communications support.
They identified 28 features that the apps had, and, among these features, some of the most prevalent ones allowed for reports to be generated, logged screen time, or blocked apps. 
From their analysis, they concluded that the features an app has do make a difference. 
The features that supported the parent-child relationship seemed to be approved of by both children and parents. 
These features generally presented tools for parents to use but didn't overwhelm the parents nor did they take away the role of the parent in the parent-child relationship. 
These good features also provided children with information about the boundaries that were set. 
Therefore, the model suggested by Hashish et al. seems to address the problems noted by prior work of a lack of communication \cite{ghosh2018matter,blackwell2016managing,wang2021protection}.

The work of Altarturi et al. examined different methods for content filtering and listed their strengths and weaknesses \cite{altarturi2020preliminary}. 
They first examine methods that are content-based web filters. 
After this, they transition to examining textual and visual methods. 
For each method, they list relevant techniques and notes about the different methods.


\subsection{Security Advice}
There is an increasing body of literature relating to the topic of security advice and the adoption of security advice \cite{boyd2021understanding,busse2019replication,fagan2016why,ion2015no,redmiles2016advice,redmiles2016census,redmiles2020comprehensive,reeder2017152}. 

The work of Redmiles et al. rated 374 pieces of advice across three axes: comprehensible, actionable, effective~\cite{redmiles2020comprehensive}.
They found that it was hard to prioritize advice, citing the example of experts classifying 118 behaviors as being part of the ``top 5'' things users should do, leaving the real burden of prioritization to the end-users~\cite{redmiles2020comprehensive}.
This is consistent with other work, like that of Reeder et al. who found that a group of 231 experts produced a list of 152 pieces of advice that should be in the ``top 3'' things to know~\cite{reeder2017152}.
Redmiles et al. also conducted 25 semi-structured interviews with a diverse pool of users to better understand where and why users take security advice~\cite{redmiles2016advice}.
They found that users tended to consider the trustworthiness of the source of advice when considering whether to follow digital advice or not, and users would typically reject advice if it seemed like there was too much promotional material in the advice~\cite{redmiles2016advice}.
In another survey of 526 users, Redmiles et al. also found that there is a difference in advice sources for users depending on their socioeconomic status and their skill levels~\cite{redmiles2016census}. 

Boyd et al. found a similar problem with prioritizing and following advice when they examined safety guides given to Black Lives Matter (BLM) protestors in spring 2020~\cite{boyd2021understanding}.
They found that only about half the guidelines explained why one should follow the advice, and only a little over a quarter of the guidelines explained how to follow the advice, leading to common pieces of advice not being followed or understood~\cite{boyd2021understanding}.

On the topic of why users might not be following advice, Fagan et al. investigated user motivation for following or not following security advice~\cite{fagan2016why}.
They found that there are gaps in the perception of users who follow common security advice and of users who do not, which may explain why some users choose not to adopt security advice~\cite{fagan2016why}.
They also found that the self-reported benefit from users, whether they chose to follow the advice or not, is higher than what the other group of users estimated the benefit of this side to be~\cite{fagan2016why}.
Also in the vein of user perceptions, Fagan et al. found that users prioritized individual concerns over social concerns when considering security advice~\cite{fagan2016why}.
Related to different perceptions of security advice, a study by Ion et al.~\cite{ion2015no} and a replication study by Busse et al.~\cite{busse2019replication} found that expert and non-expert users had significantly different priorities and habits when it came to security advice. 

\emph{We aim to add to the growing literature surrounding security advice by analyzing the quality of security advice given on TikTok and YouTube, and by examining the implications of using these platforms as a source of security advice.}

%% file: method.tex
\section{Methodology}\label{metho}

In our research, we investigated security advice found on the video-based social media platforms of YouTube and TikTok.
Our primary goal was to answer the following questions about this content:
(i) what advice is being provided to parents and children,
(ii) is the information presented accurate and actionable,
and (iii) is there a balanced discussion of the competing interests of each party?

To this end, we conduct 33 search queries on the topics of content filtering and circumvention.
These searches were conducted in March 2023 and resulted in 839 videos we tagged for analysis.
Upon analysis, only 399 of the videos contained content related to content filtering, device monitoring, or circumvention techniques.
For these 399 videos, we code them based on the target audience, topical content, accuracy, actionability, and how they discuss the interplay between parental and child rights.

This study didn't need approval from our Institutional Review Board (IRB) because it involved only publicly available data and didn't involve any interventions with human subjects.

\subsection{Search Query Selection}

We used a four-step approach for selecting the search queries used in our study:

\textbf{Step 1.} We had two researchers search on Google and Reddit for discussions of content filtering, device monitoring, and circumvention techniques.
These researchers were looking for conversations from both parents and children.
We do not analyze these conversations in this paper but do make use of the topics identified in these conversations in Step 3.

\textbf{Step 2.} Three researchers searched for and watched 230 videos on YouTube and TikTok discussing content filtering, device monitoring, and circumvention techniques.
This process did not have a formalized set of search queries and was exploratory in nature.
We do not analyze these videos in this paper but do make use of the topics identified in these videos in Step 3.

\textbf{Step 3.} We gathered a team of five researchers to select the queries to be used in our study.
This team included one researcher from Step 2, but not the other two from Steps 1 or 2.
This team was composed of a mix of genders (three male, two female), ages (18--37), ethnicities (Black, Middle Eastern, South East Asian, White), marital statuses, and number of children.
While we do not claim that this team fully represents all parents and children, we do believe that having a diverse team aided in the selection of search queries.

\begin{table*}
	\centering
	\begin{tabularx}{\textwidth}{XX}
		\textbf{Parent} & \textbf{Child} \\
		\toprule
		
		How to protect my child online & How do I unblock my device from the wifi \\
		How to secure the Internet & How do I access TikTok past bedtime \\
		How to block social media & How do I access Instagram past bedtime \\
		How to block inappropriate content & How do I access Facebook past bedtime \\
		How to block porn & How do I keep my parents from monitoring my phone \\
		How to stop kids from chatting with strangers online & How do I get around internet filters \\
		How to protect kids in online gaming & How do I access TikTok on school wifi \\
		How to disconnect devices at night & How do I access Instagram on school wifi \\
		How to disconnect devices at dinner & How do I access Facebook on school wifi \\
		How to setup OpenVPN & How do I use a VPN \\
		& How can I access my neighbor’s wifi \\
		& How can I keep my parents from restricting what I watch \\
		& How can I unlock my school Chromebook \\
		& How to get around screen time limits \\
		& How do I hide stuff on my phone from my parents \\
		& How do I hide apps on my phone \\
		\bottomrule
	\end{tabularx}
	
	\caption{Search queries used to select videos}
	\label{tab:queries}
\end{table*}

This team of five researchers began by reviewing the topics identified by the first two researchers and discussing their own experiences with these topics.
Through rigorous discussion, the team ultimately agreed on 26 search queries to use---10 for parents and 16 for children---that they felt represented queries that a parent or child would be likely to search for.
When selecting the queries, the team focused on natural wording ---i.e., wording that represented how a user would actually type the query.
This set of queries is given in Table~\ref{tab:queries}.

\begin{table}
	\centering
	\begin{tabularx}{\columnwidth}{X}
		\textbf{Parent} \\
		\toprule
		
		How to set up DNS filters \\
		How to set up content filters \\
		How to limit social media time \\
		How to see hidden apps on phone \\
		Is my internet secure \\
		Should I limit my kid's internet \\
		Should I monitor my kids online \\
		\bottomrule
	\end{tabularx}
	
	\caption{Supplemental queries used to reach saturation}
	\label{tab:supplemental-queries}
\end{table}

Fourth, after coding all the videos returned from the initial set of 26 search queries, the team of five researchers met again to discuss whether they felt saturation had been met for the topics identified in those videos.
Here, they defined saturation as meaning that any topics discussed in the analyzed videos had been covered by at least five videos.
The team agreed that for children saturation had been met.
However, they felt that for parents, several topics had come up where saturation was not met.
In this situation, the team felt that parents would make additional searches to learn about these videos, so seven additional search queries were added for parents.
These queries are listed in Table~\ref{tab:supplemental-queries}.

\subsection{Video collection}

We executed each search query once on YouTube and once on TikTok.
Search queries were executed using the official YouTube API and the unofficial TikAPI for TikTok, with the top 200 results being stored for each platform.
For each video, we stored not only the video but also metadata about the video, such as its author, description, and engagement (i.e. views, likes, and comments).

To account for the influence of personalized algorithms on search query outcomes, we ran the search queries for parents and children using separate accounts.
We also explored whether the account's reported user age impacted search results by creating accounts with a range of ages and running the queries on those accounts.
After comparing the results from these different accounts, we found no noteworthy differences, so we only evaluated the videos retrieved from a single parent (age 40) and a single child account (age 17).
These accounts were new accounts that had not been used in the exploratory study.

\subsection{Video Analysis}

Videos from the search queries were watched and analyzed by four researchers, with two researchers being assigned the parent-related videos and two researchers the child-related videos.
Each pair of researchers worked together, viewing and coding videos as a team.
We do not report inter-rater reliability, as codes were only assigned to a video after complete agreement was reached.

For each search query, the relevant pair of coders would begin analyzing videos from the most to the least relevant (as indicated by their order in the search results).
Coders would continue coding videos from that query until they felt they had reached saturation, defined as watching eight videos in a row in which either all videos were no longer relevant to the search term or until the videos were revealing no new information about the search topic.
At a minimum, coders would code at least 10 videos from each search term.
In total, the researchers watched 839 videos. 

Videos were coded using a codebook (described below).
Before beginning coding, the codebook creators and the coders met together for training.
In this training, the contents of the codebook were described, examples were given for each code, and definitions for terms were clarified.
Coders were free to ask questions throughout this process.
At the end of this training, coders each coded ten practice videos to ensure that they fully understood how to use the codebook and that code assignment was consistent between the coders.
This exercise succeeded, so after discussing their experiences with each other, the coding of videos began.

\subsubsection{Codebook}

Questions in the codebook were close-ended.
Coders did have the ability to note if there was meaningful information in the video that didn't fit into the provided coding, but this functionality was not ultimately needed as the codes in the codebook turned out to be sufficient to describe the videos.
The codebook is given in full in Appendix~\ref{appx:codebook}, with a summary of its contents below.

First, coders indicated whether after watching the video it was relevant to the search query.
Here we use a very broad definition of relevance---i.e., if it touched in any degree on content filtering, device monitoring, or circumvention techniques, it was considered relevant.
If the video was not relevant, nothing more about the video was coded.
In total, 399 videos were coded as relevant.

Second, the codebook asked whether the video was professionally produced, whether it was trying to be funny or meme-like, whether it had been sponsored by a company, what information was contained in the video's description, and the target audience of the video (parent or child).

Third, the codebook asked what topics were discussed in the video.
It also asked what types of devices were discussed in the video.
The codebook also asked what stance, if any, the video took regarding parental vs. children rights and whether the video gave reasons for why parents might need content filtering or device monitoring.

Fourth, the codebook asked how accurate, comprehensive, and actionable the videos were~\cite{redmiles2020comprehensive}.
A video was considered accurate if the information it provides is correct and up-to-date.
A video was comprehensive if it explained all necessary context regarding a technology.
A video was actionable if it presented explicit steps that could be taken by the viewer as opposed to only being an informational discussion of a subject.
For each of these three items, the options were ``Yes,'' ``Somewhat,'' and ``No''~\cite{jacoby1971three}.
If the answer to any of these three items was not ``Yes'', then coders would indicate what the problem was (open-ended prompt).
Finally, the codebook asked whether the video's title accurately described its contents (``yes''/``no'').

\subsubsection{Codebook Development}

To develop our codebook, three researchers first conducted an exploratory study of content filtering videos on YouTube and TikTok.\footnote{This study was also used as Step 2 of our query selection process.}
Between September 2022 and November 2022, these researchers gathered and analyzed 230 videos from YouTube and TikTok.
Of these, 115 focused on educating parents and guiding them in setting up parental controls, while the remaining 115 focused on techniques for children to use to bypass content filtering.
As this was an exploratory study, the selection of the search queries and videos was left up to the discretion of the researchers, with the researchers continuing to search for videos until they had reached saturation.

Throughout the process of gathering videos, the researchers met together and analyzed the videos using open coding and the constant comparative method~\cite{glaser1965constant}.
Results from this coding were used to inform the selection of further videos.

At the conclusion of this exploratory study, three researchers used the data from the exploratory study to create the codebook for the main study.
These three included one of the researchers who conducted the exploratory study, and all three members who created the codebook participated in query selection.
To create the codebook, the three researchers first drafted a codebook that covered all codes created from the exploratory study.
Then, through discussion and repeated revision, they continued to refine the codebook, combining related codes and removing unusual codes, until they arrived at a codebook they felt could have accurately coded all the videos from the exploratory study.
This codebook was then used in the final study.
The experience gained from reviewing the exploratory study results and developing the codebook was also helpful when these researchers trained the coders.

\subsection{Limitations}

While we've aimed to minimize potential biases in our approach, it's important to acknowledge a few factors that could affect the ecological validity of our data.

First, since we stopped evaluating videos when reaching saturation, it is possible that some later videos may have revealed new topics.
While we believe it is unlikely that most users would continue so far down a list of irrelevant videos to get to these hidden gems, it is still something worth keeping in mind when considering our results.

Second, while we took steps to mitigate the impact of personalized algorithms on search results, as these algorithms are opaque, we can't guarantee they didn't manage to personalize the results to some extent.
For example, they may personalize videos based on the locality of the IP block we used.

Third, the videos collected represent a snapshot in time.
The YouTube and TikTok algorithms are highly dynamic and opaque meaning that the videos returned by search algorithms could change rapidly over time.
While we gathered a large number of videos from a wide range of search queries, future research will be needed to see how consistent our results are over time.

%% file: result.tex
\section{Results}\label{res}

\begin{table}
	\centering
	\begin{tabular}{l|rr}
		& \multicolumn{1}{c}{\shortstack{Total\\videos}} & \multicolumn{1}{c}{\shortstack{Relevant\\Videos}}\\
		\midrule
		YouTube & 489 & 274 (56\%) \\
		TikTok & 350 & 125 (36\%) \\ \midrule
		
		Parent & --- & 200 \\
		Children & --- & 199 \\ \bottomrule
	\end{tabular}
	\caption{Summary of videos by platform and by audience}
	\label{tab:video_breakdown}
\end{table}

We analyzed and coded a total of 839 videos.
Of those, only 399 (48\%) ended up being relevant to content filtering, device monitoring, or circumvention techniques.
Table~\ref{tab:video_breakdown} breaks down video counts by platform and target audience.

\subsection{Engagement}

\begin{table*}
	\centering
	\begin{tabular}{l|rrr|rrr|rrr}
		& \multicolumn{3}{c}{Views} & \multicolumn{3}{c}{Likes} & \multicolumn{3}{c}{Comments} \\
		& 
		\multicolumn{1}{c}{Min} & \multicolumn{1}{c}{Median} & \multicolumn{1}{c|}{Max} & 
		\multicolumn{1}{c}{Min} & \multicolumn{1}{c}{Median} & \multicolumn{1}{c|}{Max} & 
		\multicolumn{1}{c}{Min} & \multicolumn{1}{c}{Median} & \multicolumn{1}{c}{Max} \\ \midrule
		
		Parents
		& 0 & 5,684 & 11,058,350
		& 0 & 68 & 100,344
		& 0 & 9 & 24,569 \\
		
		Children
		& 0 & 50,442 & 12,086,228
		& 0 & 443 & 198,535
		& 0 & 49 & 10,579 \\ \bottomrule
	\end{tabular}
	\caption{Video engagement by audience}
	\label{tab:interactions}
\end{table*}

Looking at the engagement these videos generate (see Table~\ref{tab:interactions}), it is clear that children interact with videos around content filtering on YouTube and TikTok much more than do parents. 
Still, while it is less common, parents are interacting with these videos, with some having millions of views, tens of thousands of likes, and tens of thousands of comments.
So, while parents interact with these video-sharing platforms less than children, they are still a common information source for some parents.
In fact, the videos with the largest number of comments were targeted at parents.

\subsection{Video Topics}

\begin{figure*}
	\centering
	\includegraphics[width=.8\textwidth]{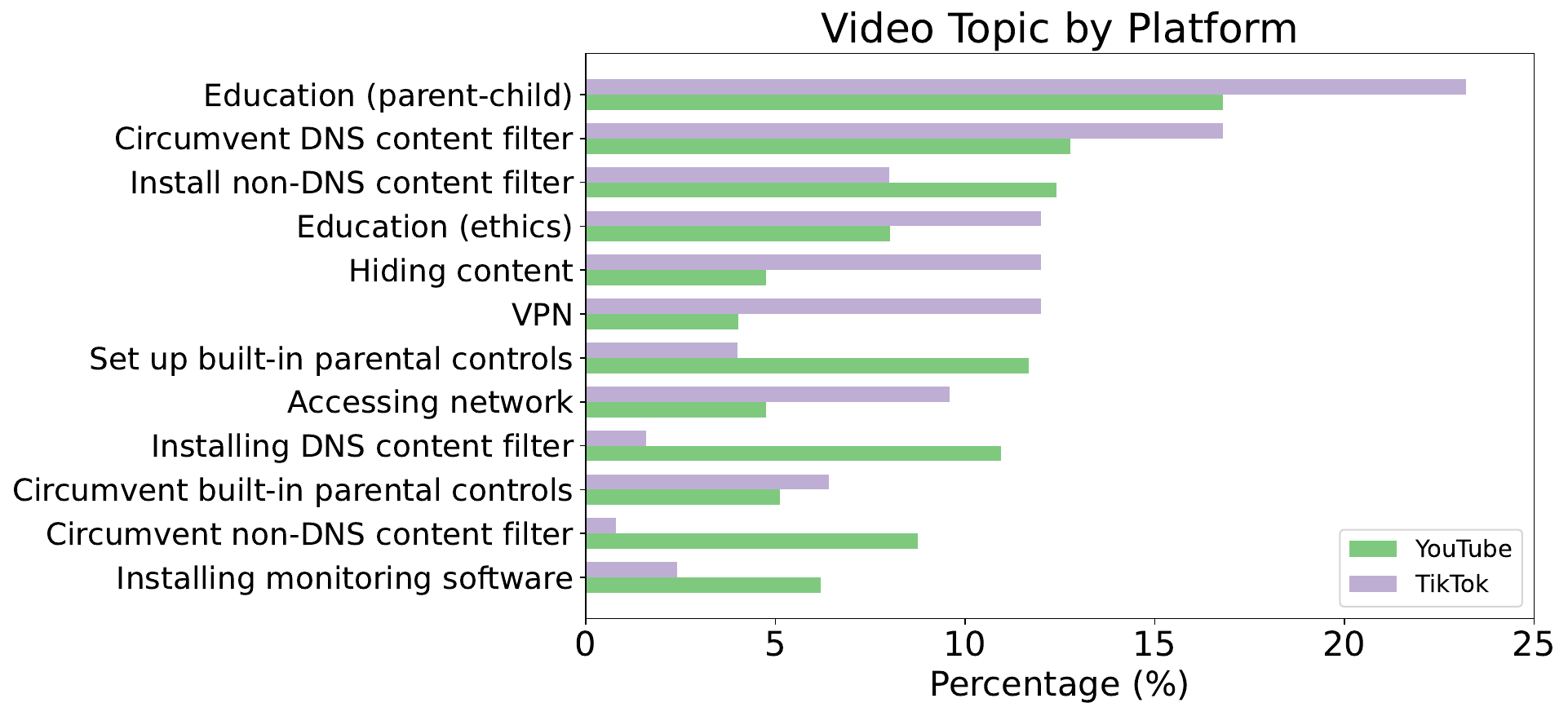}
	\caption{Video topics by platform}
	\label{fig:video_topic_platform}
\end{figure*}

\begin{figure*}
	\centering
	\includegraphics[width=.8\textwidth]{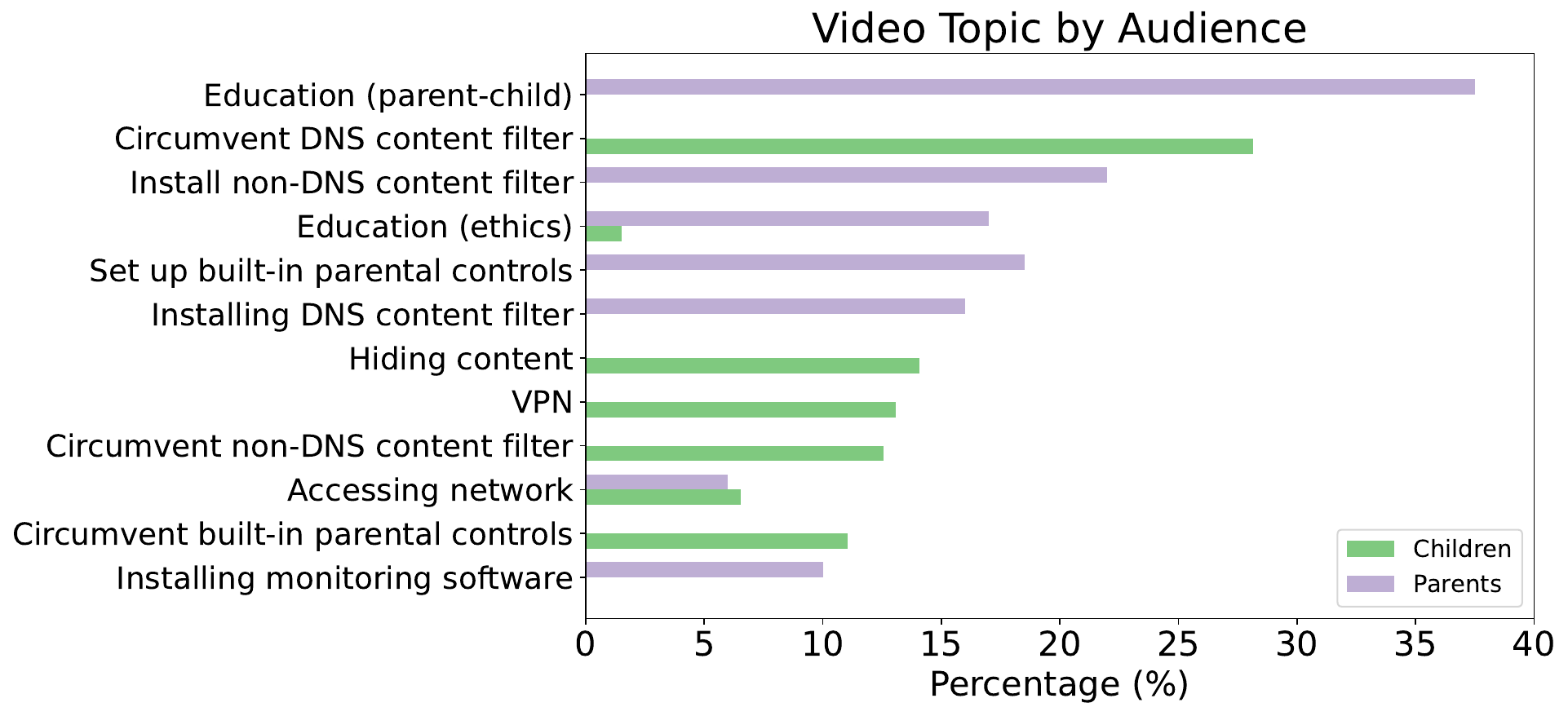}
	\caption{Video topics by audience}
	\label{fig:video_topic_audience}
\end{figure*}

Figures~\ref{fig:video_topic_platform} and \ref{fig:video_topic_audience} list the most common video topics split by platform and audience, respectively.
One interesting trend in both graphs is that videos commonly discuss circumventing DNS content filters, but installing DNS content filters is a less common topic.
On the flip side, while installing non-DNS content filters is a common topic, circumventing non-DNS content filters is a less popular topic.
This is an interesting disparity and one that could be examined more in future work.
For example, it might be the case that while people want to install non-DNS content filters, doing so requires too much work, leading them to adopt DNS content filters, explaining why circumvention of the latter type of filters is discussed in more videos.

Looking at platform-specific differences, we see that TikTok hosts a higher number of videos centered on general online safety education, which stands out as the most frequently discussed topic.
Additionally, YouTube excels in terms of videos addressing the establishment of parental controls, whereas TikTok leans towards content related to circumventing these controls.
In general, setting up parental controls prevails on YouTube, while strategies to bypass them prevail on TikTok.

Looking a audience-specific differences, we find minimal overlap between the topics parents and children see in videos.
This is to be expected as the goals of both groups are quite different.
The sole exception to this is the topic of network access which is viewed by both audiences.

\subsection{Education and ethics}
One extremely interesting difference between topics for parents and children relates to general education about online threats and discussion of ethics.
As shown in Figure~\ref{fig:video_topic_audience}, parent-oriented videos often provide general background about Internet safety for both parents and children.
However, this topic is completely missing for children.
So, while they are learning how to circumvent parental controls, there is no content making them aware of the dangers that this might bring.

\begin{table}
	\centering
	\begin{tabular}{l|rrr}
		& \multicolumn{1}{c}{\shortstack{Parent\\rights}} & \multicolumn{1}{c}{\shortstack{Child\\rights}} & \multicolumn{1}{c}{Nuanced} \\ \midrule
		
		YouTube & 6 & 8 & 10 \\
		TikTok & 7 & 8 & 0 \\ \midrule
		
		Parent & 13 & 13 & 10 \\ 
		Child & 0 & 3 & 0 \\ \bottomrule
	\end{tabular}
	\caption{Summary of ethical stance by platform and by audience}
	\label{tab:ethics}
\end{table}

Even more interesting, we examined whether videos discuss the ethics of parental filtering and monitoring.
Table~\ref{tab:ethics} summarizes our findings.
While this topic is not discussed that often ($n=39$,~10\%), it is nearly only ever discussed in parent-oriented videos.
In a third of these videos, parents are presented with the ethical reasons parents should be able to monitor their kids' devices.
In a third of the videos, parents are presented with ethical concerns about monitoring kids' devices, with a recommendation that they avoid doing so.
In the final third of the videos, parents are told how this is a nuanced issue that deserves careful consideration.
We believe it is great to see that parents are being provided with diverse viewpoints on this topic.

In stark contrast, only three child-oriented videos even discuss the ethics of monitoring, with all three videos taking an anti-parent stance.
Taken in light of the lack of content educating children about online threats generally, we find this situation to be potentially harmful.
While there can be many abusive uses of monitoring and filtering technology, it can also be used to protect children.
However, children are not being informed of this purpose, giving them a one-sided view of this issue, and inhibiting their ability to make informed choices.

\subsection{Quality: Accurate, Comprehensive, and Actionable}

\begin{figure*}
	\includegraphics[width=.8\textwidth]{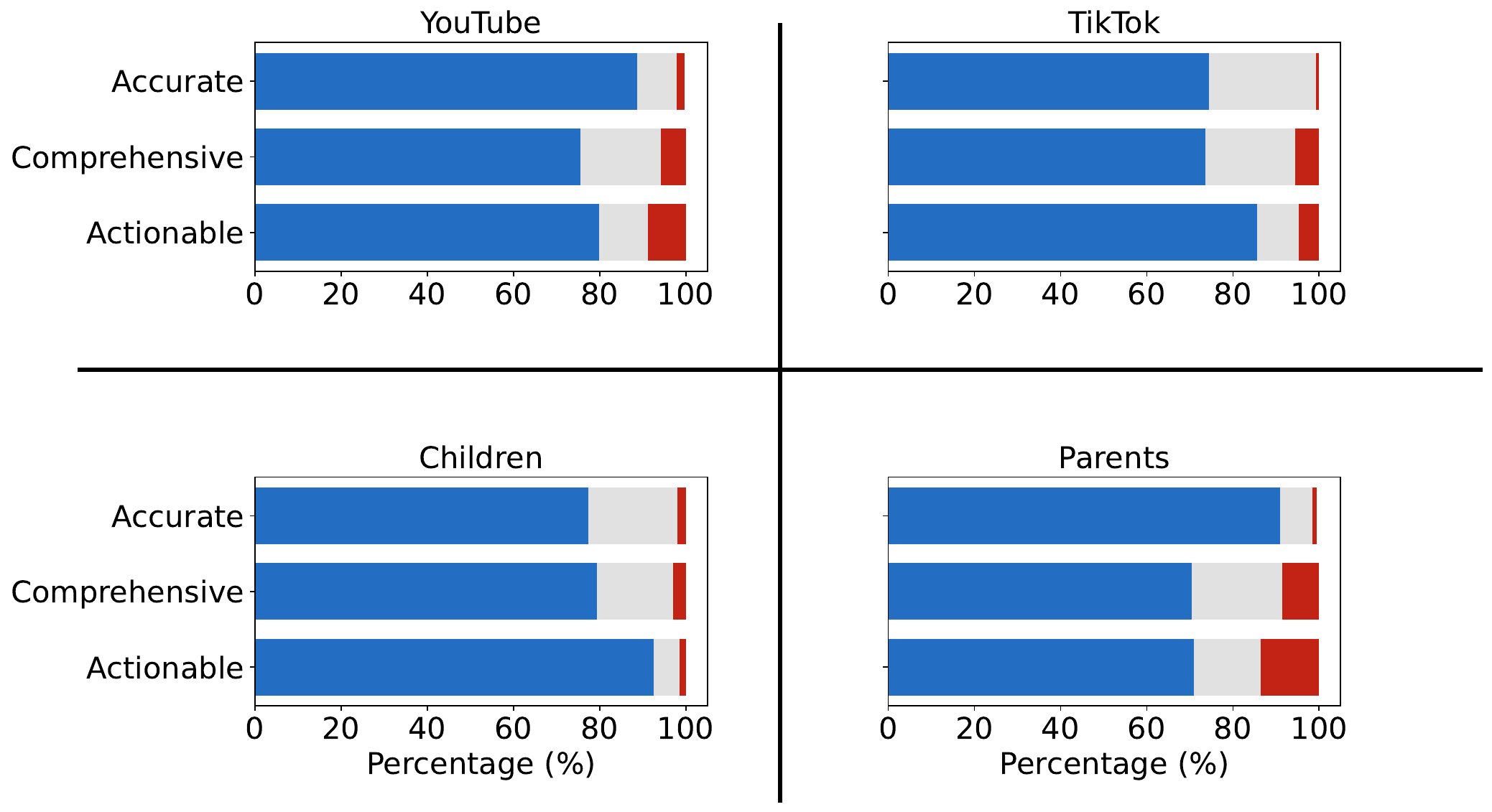}
	\includegraphics[width=.55\textwidth]{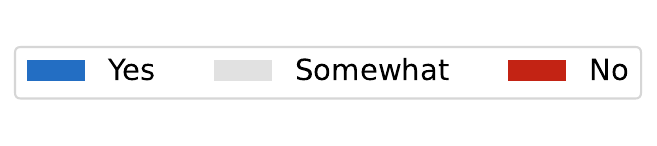}
	\caption{Video quality ratings by platform and audience}
	\label{fig:aca}
\end{figure*}

Figure~\ref{fig:aca} shows the accuracy, comprehensiveness, and actionability of the videos we coded.
Overall, 337 videos (84\%) of videos were accurate, 299 (75\%) were comprehensive, and 326 (82\%) were actionable.
While this is far from perfect, we were still surprised at the overall quality of the videos.

Breaking down these metrics by platform, we see that YouTube videos ($n=243$,~89\%) are more likely to be accurate than TikTok videos ($n=93$,~74\%), with the difference being statistically significant ($\chi^2(2)=17.68$, $p<0.001$).
This suggests that YouTube might be a much better source for security advice and is something that future research should examine more in-depth.

In contrast, there is no difference in comprehensiveness ($\chi^2(2)=0.26$, $p=.88$) or actionability ($\chi^2(2)=2.3616$, $ p=.31$) for videos based on platform.
This result was uprising to us as we expected the short-form nature of TikTok videos to significantly impact their ability to be comprehensive and actionable, but this didn't turn out to be the case.

Comparing videos based on audience, we see that videos targeting children are more likely to be actionable ($n=184$,~92\%) than they are for parents ($n=142$,~71\%) ($\chi^2(2)=33.00$, $p=<0.001$).
In contrast, the accuracy of videos for parents ($n=182$,~91\%) is higher than that of videos for children ($n=154$,~77\%) ($\chi^2(2)=15.07$, $p<0.001$), as is the case for comprehensiveness ($\chi^2(2)=6.86$, $p<0.05$), though in that case the effect size is small.
We find these results somewhat troubling.
In particular, in the case of children, they are more likely to receive actionable steps that do not address their problems and may lead to more issues. 

\subsection{Video Style}

\begin{table}
	\centering
	\begin{tabular}{l|rrrr}
		& \multicolumn{1}{c}{Serious} & \multicolumn{1}{c}{Professional} & \multicolumn{1}{c}{Sponsored} \\ \midrule
		
		YouTube & 264 & 168 & 17 \\
		TikTok & 119 & 74 & 5 \\ \midrule
		
		Parent & 200 & 174 & 17 \\ 
		Child & 196 & 73 & 8 \\ \bottomrule
	\end{tabular}
	\caption{Summary of video styling by platform and by audience}
	\label{tab:style}
\end{table}

Table~\ref{tab:style} summarizes the styling of the videos we analyzed.
We examined videos to see if they used a serious or comedic tone, finding that videos nearly universally ($n=396$,~99\%) took a serious tone.
Nearly two-thirds appeared to be professionally produced ($n=247$,~62\%).
Interestingly, very few ($n=25$,~6\%) sponsored any products.

\subsection{Devices}

\begin{figure*}
	\centering
	\includegraphics[width=.65\textwidth]{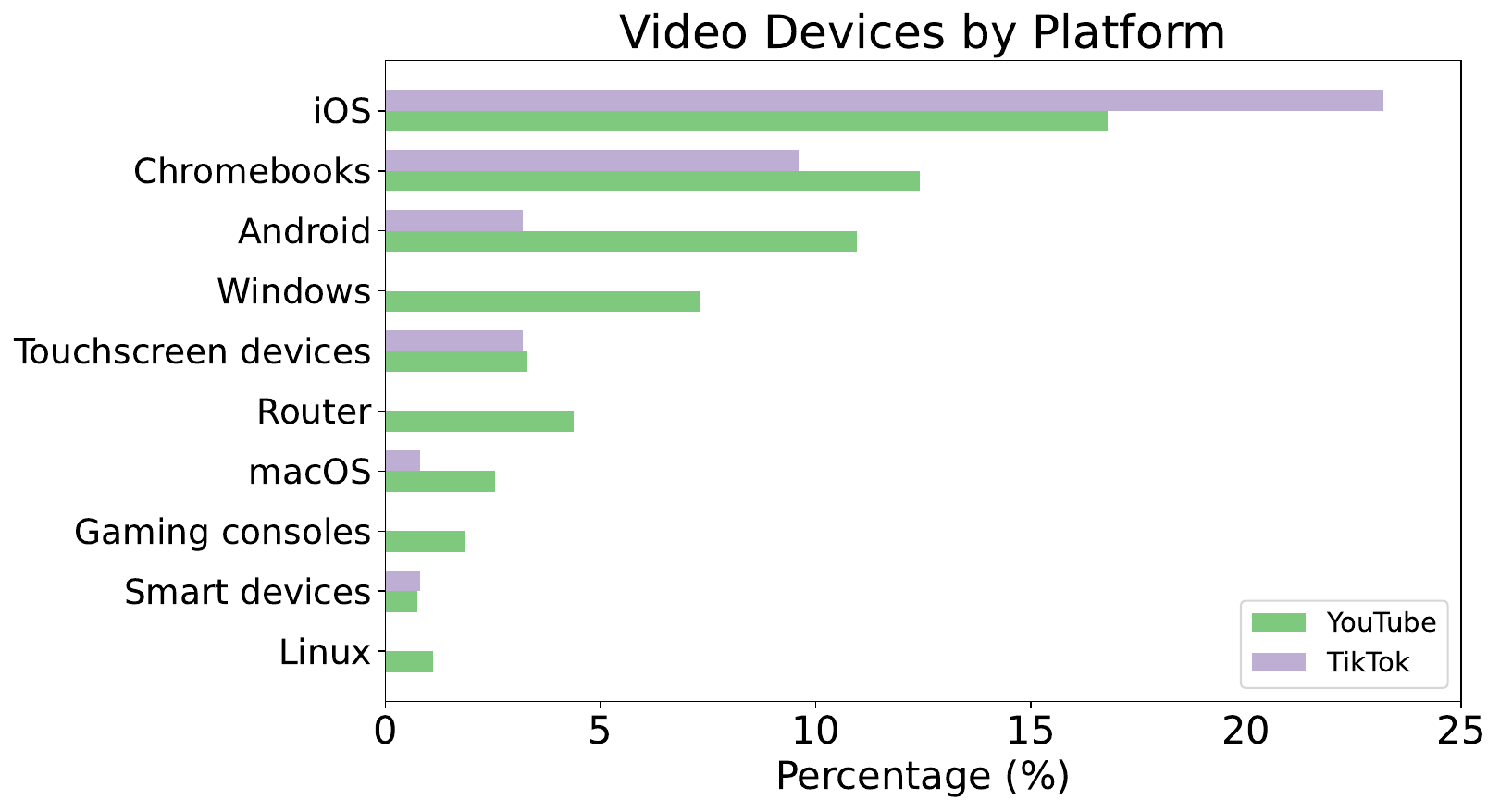}
	\caption{Devices mentioned by platform}\label{fig:video_device_platform}
\end{figure*}

\begin{figure*}
	\centering
	\includegraphics[width=.65\textwidth]{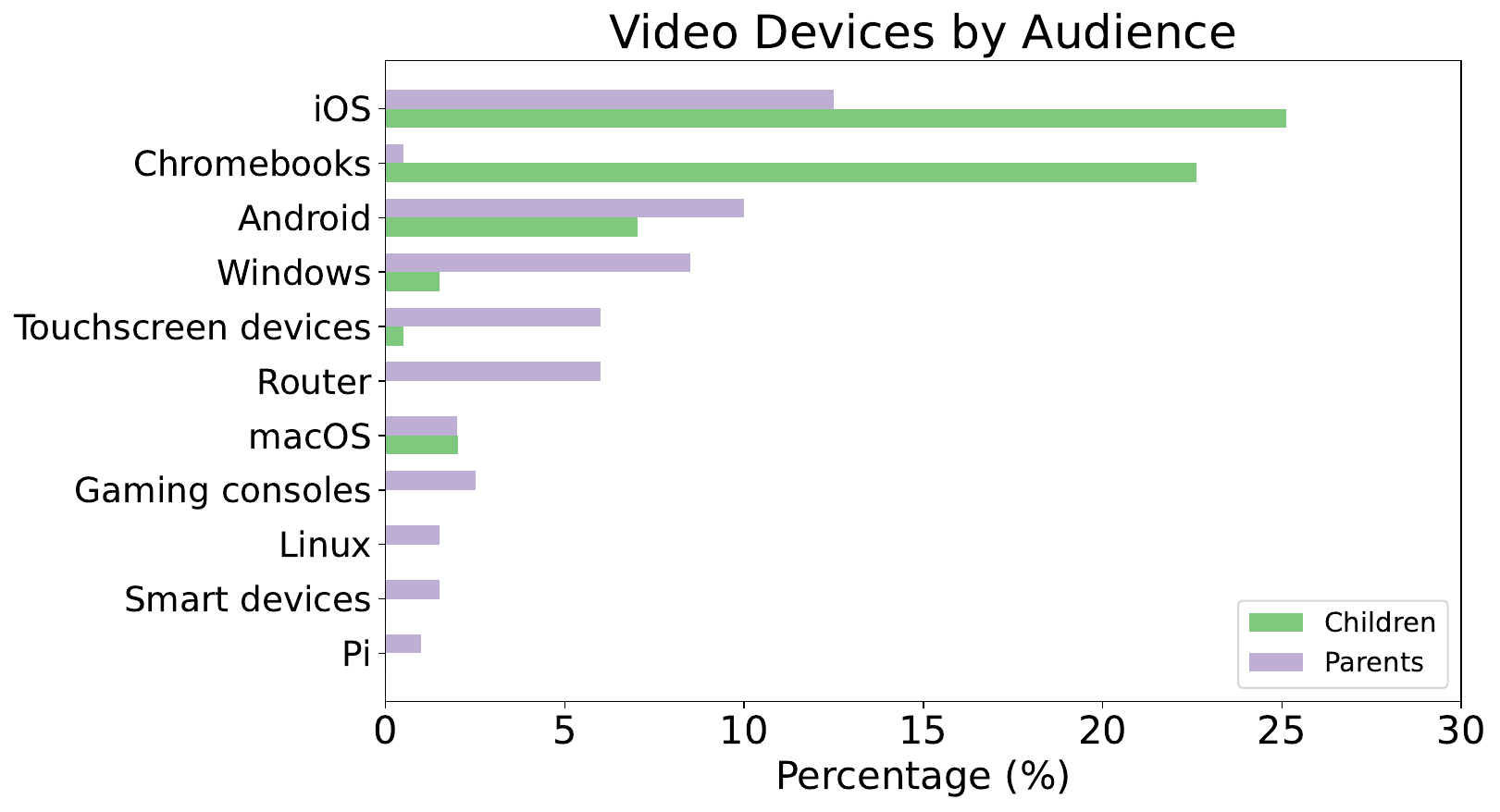}
	\caption{Devices mentioned by audience}\label{fig:video_device_audience}
\end{figure*}

Figure~\ref{fig:video_device_platform} provides insights into the devices mentioned within the videos, with data segmented by their originating platform.
We have included only those devices that appeared in at least 1\% of the videos.
It is worth noting that not all videos made specific device mentions.
Notably, TikTok videos tend to mention iOS devices more frequently compared to non-iOS counterparts, such as Android or Windows devices.
This data suggests that TikTok primarily focuses on devices commonly used in everyday life, notably iPads and Chromebooks. 
The prevalence of Chromebooks and iOS devices, like iPads, on TikTok, can be attributed to their widespread use in educational settings~\cite{ahlfeld2017device,kaur2020post,henderson2012ipad,alyahya2012ipads}.
In contrast, YouTube mentions more specific devices more frequently than TikTok.

Figure~\ref{fig:video_device_audience}, which groups device data based on the intended audience of the video, reinforces these findings.
Chromebooks, for instance, are predominantly mentioned in videos targeting children, reflecting their extensive use in educational contexts.
Additionally, over half of the iOS device mentions are associated with videos intended for children, further highlighting the role of iPads and similar devices in educational settings.
In contrast, devices such as routers, gaming consoles, or those running on Linux are almost exclusively mentioned in videos aimed at parents, likely aligning with their role in managing household technology and network infrastructure.

\subsection{Content Filtering and Bypassing Motivations}

\begin{table*}
	\centering
		\begin{tabular}{l|rr}
			& \multicolumn{1}{c}{YouTube} & \multicolumn{1}{c}{TikTok}\\ \midrule
			
			Inappropriate materials to kids & 53  & 16 \\
			Protect kids from malicious individuals & 36  & 18 \\
			Social media is dangerous to kids & 29  & 13 \\
			Internet is distracting & 32  & 7 \\
			Malicious software that kids can accidentally download & 9  &  0 \\ \bottomrule
		\end{tabular}
	\caption{Parents' motivations for adopting content filtering}
	\label{tab:reasons}
\end{table*}

\begin{table*}
	\centering
		\begin{tabular}{l|rr}
			& \multicolumn{1}{c}{YouTube} & \multicolumn{1}{c}{TikTok}\\ \midrule
			
			Access whatever webpages the child wants & 63  & 29 \\
			Gain unauthorized access to network & 10  & 17 \\
			Getting around time-based restrictions on internet usage & 13  & 4\\
			Strict parents & 1  & 13 \\
			Gain increased access to social media & 9  &  2 \\ \bottomrule
		\end{tabular}
	
	\caption{Children's motivations for circumventing content filtering}
	\label{tab:circumvent}
\end{table*}

Table~\ref{tab:reasons} outlines the primary motivations presented in videos regarding why parents should implement content filters for their children's online interactions.
Similarly, Table~\ref{tab:circumvent} gives the most common motivations for children to bypass content filters or device restrictions.
While none of these motivations are surprising, it is still interesting to observe their relative frequency in these video information sources.

%% file: goals.tex
\section{Goals and Techniques}

In this section, inspired by the work of Wei et al.~\cite{wei2022anti}, we detail the different goals presented in the videos for the context of setting up content filtering and circumventing them, drawing insights from our comprehensive analysis. It's worth mentioning that any comments mentioned in this section are derived from an analysis of the top 10 most relevant comments, as decided by the platform's sorting algorithm, on the corresponding video.


\subsection{Content Filtering}

\subsubsection{Parental Concerns}

Parents express various concerns about their children's online activities, motivating the implementation of content filtering measures. The primary goals include:

\paragraph{Inappropriate Content Protection}

Parents aim to shield their children from accessing inappropriate materials on the internet. Our study found that 53 YouTube videos and 16 TikTok videos specifically addressed this concern.

\paragraph{Securing Children from Online Threats}

Approximately 36 YouTube and 18 TikTok videos emphasized the need to protect children from malicious individuals online.

\paragraph{Managing Internet Distractions}

Parents videos including 32 on YouTube and 7 on TikTok, seek to control and limit their children's time spent online, viewing the internet as a potential distraction.

\subsubsection{Techniques Employed}

To achieve these goals, parents utilize various techniques, as highlighted in the analyzed videos:

\paragraph{Content Filters Implementation}

Creators demonstrated the use of content filters, including DNS filters and parental control apps, as effective tools for restricting access to inappropriate content. In particular, one of the most watched videos on YouTube (1.9k views) focused on explaining what DNS filters are in a very simple non-technical way and then showed the exact steps to set up a DNS filter. The comments on this video showed interest and enthusiasm about the creator's method and way of explanation, with comments such as \textit{``You explain things so clearly, thank you''} and \textit{``This is so helpful!''} Parents started raising further questions commenting on this video as \textit{``How you add extra layer of protection?? Plz tell. Thanks in advance.''} and \textit{``What are some router suggestions. We have 2 streaming tvs. Occasionally play a video game. 3 cell phones and use Wi-Fi calling. Our house is around 3500sqft.''}

\paragraph{Location Tracking Apps}

Concerned about their children's safety, parents employ location-tracking apps like ``Find My'' to monitor their children's whereabouts, as shown in videos with high engagement. A video describing how to use the ``Find My'' app had 279k views at the time of the study. It describes how to track the location of the kids when they are away from home as well as how to set up notifications for when kids reach home. Parents seemed satisfied with using this technique as it is not costly and easy to use, according to some of the comments: \textit{``I was spending \$69/yr for a program (I won't name) primarily for these features! I use it for my 13 yr old. Ty so much! Very clear, direct, with easy to follow instructions''}, and \textit{``I use this app for my 18 year old son with autism who lives on a college campus. I can see when he goes to class and when he is still in his dorm at 10am possibly sleeping. Works great for when he goes out of town with band or baseball team.''}. Other suggested apps with fewer interactions were ``Life360'' and also ``Bark''.

\subsection{Circumventing Content Filtering}

\subsubsection{Children's Motivations}

Children, on the other hand, are motivated to bypass content filters and restrictions for various reasons, as outlined in the following goals. In the videos that we collected in the context of bypassing content filters, almost all of the videos explicitly started by mentioning malicious goals such as: ``HOW TO BYPASS Parental Control Settings! NEW | Working 2022'', ``how to unpause wifi your parents blocked'', ``HOW TO BYPASS ANY WEB FILTER!'' and ``How to Bypass School Internet Filters \& Restrictions in 5 simple steps!''

\paragraph{Unrestricted Access}

Videos targetting children, including 63 on YouTube and 29 on TikTok, teach users how to gain unrestricted access to webpages without content filtering constraints.

\paragraph{Avoiding Time-Based Restrictions}

Approximately 13 YouTube and 4 TikTok videos suggest that children often attempt to circumvent time-based restrictions on internet usage imposed by their parents.

\paragraph{Circumventing Strict Parental Controls}

In cases where parents enforce strict controls, children (13 on TikTok) are driven to find ways to regain control over their internet access.

\subsubsection{Techniques Employed}

Children employ a variety of techniques to circumvent content filters, showcasing creativity and adaptability:

\paragraph{VPN and Proxy Usage}

Virtual Private Networks (VPNs) and proxies are popular among children, as demonstrated in numerous videos on both platforms. These methods provide a straightforward way to bypass network filters. 14 videos on YouTube and TikTok with over 3 million views focused on showing kids how to install and activate VPNs. One of the videos called ``How To Bypass WiFi Restrictions!'' which talks about how to use a VPN to bypass wifi restrictions, was just under 2 minutes and had 111K views at the time of the study and the audiences were mostly kids, according to the comments on this video. However, this large number of views doesn't always indicate that the method is working as confirmed by one of the comments: \textit{``blocked...''}.

\paragraph{Device Resets}

Children resort to resetting devices, particularly evident in videos providing instructions on bypassing restrictions on school-issued Chromebooks.

\paragraph{Accessing Alternate Networks}

Videos demonstrate children obtaining alternate routers or accessing nearby networks, including attempts to guess passwords, as a means to bypass content filters. Five videos demonstrated obtaining an alternate router (one video even mentioned how to get a new router for free and how to set it up!). Others were showing how to illegally get access to nearby networks (i.e. neighbors' wifi networks). Some of those videos speculate that attempting to guess the password of a home network is possible using default passwords. An easier and more logical technique proposed was using mobile data or hotspots to bypass any network filters.

\paragraph{Device-Specific Techniques}

Certain videos reveal device-specific techniques, such as changing network/MAC addresses, often exploiting features like ``private wifi address'' on iOS devices.

%% file: discussion.tex
\section{Discussion}\label{dis}

In this section, we discuss the implications of our results.

\subsection{Video-Based Social Media as a High-Quality Information Source}
\label{sec:dis:source}
Our findings show that security advice videos on YouTube and TikTok are already beginning to see high engagement.
The impact of this development is not inherently good or bad; its quality depends on the accuracy, comprehensiveness, and actionability of the information shared on these platforms.

Overall, we were surprised about the quality of videos on these platforms.
Roughly three-quarters of videos on both platforms were accurate, comprehensive, and actionable.
This means that if users watch enough videos, they will get the information that satisfies their needs.
Additionally, these videos are getting reasonable engagement, from thousands to millions of views and hundreds to thousands of likes.
This indicates that YouTube and TikTok are already, to some extent, effective platforms for security advice dissemination.

However, finding the appropriate information may not always be easy.
Prior research has shown that users often struggle to discern and prioritize the advice they receive \cite{redmiles2020comprehensive,boyd2021understanding}.
This is likely to be the case for advice found on YouTube and TikTok.
As such, we think there is room for work by security researchers and practitioners to help improve the effectiveness of these platforms as security advice sources.

First, researchers and practitioners could participate in the generation and publication of security advice videos.
These videos are highly likely to be accurate, comprehensive, and actionable, increasing the quality of videos on these platforms.
Moreover, as security advice videos on these platforms can achieve high engagement, this might provide a mechanism to share security advice based on recent research, something that has traditionally been hard to achieve.

Second, we think there needs to be research into mechanisms for more effectively helping users filter out irrelevant, inaccurate, or non-actionable advice.
Automatically determining the relevancy or inaccuracy of videos may not be a tractable problem, and as such we advocate for research into crowdsourced approaches to solving this problem.
This could include allowing experts to annotate videos or allowing users to collectively flag videos~\cite{chan2021reliability}  (which has become more difficult with the removal of dislike counts on many social media platforms).
While such approaches have traditionally involved a binary determination (good or bad), research could explore whether allowing more fine-grained ratings around accuracy, comprehensiveness, or actionability could be useful.

\subsection{Quality Issues in Children-Targeting Videos}

Our results show that videos targeting children are very actionable (92\%), but are less likely to be accurate (77\%).
These actionability numbers are encouraging---research has long shown security advice and recommendations need to be actionable~\cite{ruoti2016private,redmiles2020comprehensive}.
However, there is danger from content that is actionable but inaccurate.
Such content increases the likelihood that users will take action that could be harmful.
This is particularly concerning in the case of children, a vulnerable population that may not yet fully understand the implications of taking incorrect action.

This risk is further augmented by the fact that we did not find any videos targeted at children that explained the positive benefits of content filtering and device monitoring.
While there is clearly potential for the abuse of these technologies and justifiable reasons for children to circumvent them, such as in the case of abusive home environments, there can also be significant safety provided by these technologies.
Before circumventing these technologies, ideally, children would be informed about both the benefits and consequences of doing so, allowing them to make more informed decisions.
However, we found no such videos returned by our search queries.

In both cases, we feel that children are being underserved.
In contrast, videos targeting parents are more likely to be accurate (91\%) and often discuss the complicated ethics around parental content filtering and device monitoring.
While this content isn't perfect either, it provides parents with a more holistic view of the situation, allowing them to make informed decisions.

As such, we think there is an urgent need to both produce more content for children from trusted sources as well as provide them with more effective filtering tools.
We believe this ties into the research agenda laid out in \S\ref{sec:dis:source}.

\subsection{Difference Between YouTube and TikTok}

\subsubsection{Search Relevance}
Our analysis revealed that on TikTok, the search results quickly became less relevant, sometimes within just 5--6 videos. In contrast, YouTube consistently provided relevant results even after examining up to 25 videos in a search.

Future research should aim to explore the underlying causes of this discrepancy.
It could investigate whether TikTok's relatively younger platform age contributes to this phenomenon, possibly due to insufficient information or differing algorithms that influence search result relevance
Understanding the factors influencing the search results' relevance on these platforms could yield valuable insights into their functioning and potential areas for improvement.

\subsubsection{Content Depth}
TikTok specializes in delivering easily digestible short-form content.
TikTok videos typically offer quick overviews of various methods, making it easy for users to gather a lot of information rapidly.
These videos often include personal stories and anecdotes, fostering a stronger connection between the user and the presented problem.
However, the brevity of TikTok videos can hinder a deep understanding of technicalities and nuances.

In contrast, YouTube excels in providing in-depth, lengthy content that delves into the why behind different methods and explores various alternatives.
While this detailed approach is beneficial, it comes at the cost of longer video durations and a relative lack of personal stories, making it challenging for users to explore multiple methods efficiently. Notably, there was a noticeable drop in video quality when transitioning from technical search queries to more general ones on YouTube.

Based on these findings, we propose a strategic approach to optimize the learning experience, novice users can start by exploring different methods through TikTok's short-form content to gain familiarity with various technical concepts. Subsequently, they can transition to YouTube for in-depth learning, leveraging each platform's strengths to acquire a comprehensive understanding of security advice.

%% file: conclusion.tex
\section{Conclusion}\label{conc}

In this paper, we study the content and quality of informational videos found on the video-sharing sites YouTube and TikTok.
We focus on parent-child contexts, where parents aim to safeguard their child's online experience through content filtering and time restrictions, while children, especially teenagers, find this an invasion of their privacy and seek different methods to circumvent these restrictions.
Our research aims to provide insights into how families navigate this complex situation, seeking to offer insights that can inform discussions around privacy, security, and family dynamics in the digital age.

To this end, we collected and analyzed 839 videos from YouTube and TikTok, identifying 399 related to this topic.
We analyzed the content, tone, and styling of these videos to understand the information being provided to parents and children using these platforms to gather security advice. 
Despite the negative connotations surrounding these platforms, our analysis unveiled a reservoir of valuable security advice within video-based social media, with over three-quarters of videos providing accurate, comprehensive, and actionable information.


Within the videos, we found that content focused on bypassing restrictions tended to offer practical and straightforward guidance, whereas videos targeted at parents often comprised advertisements or general educational content. This observation underscores the advantage children, acting as potential attackers, possess in navigating the online landscape compared to parents.
In contrast, we find that parents---and not children---are the only audience receiving information about the ethics of parental monitoring and content filtering and the risks of circumventing these protections.
Neglecting these aspects may not only strain the parent-child relationship but also hinder the healthy development of these platforms as valuable sources of information. Effective communication and ethical deliberations are essential to harnessing the full potential of these platforms while safeguarding children's well-being.
As such, while YouTube and TikTok are promising avenues for security advice, there is clearly still work to be done in improving the quality of content on these platforms.

%% file: appx_codebook.tex
\section{Codebook}\label{appx:codebook}

Video ID \hrulefill \\

\noindent Relevant? \\
\begin{itemize*}
	\item Yes
	\item No
\end{itemize*} \\
\textit{Skip To: End of Survey If Relevant? = No} \\

\subsection{Overview}

\noindent Did the video appear to be professionally produced? \\
\begin{itemize*}
	\item Yes
	\item No
\end{itemize*} \\

\noindent Was the video trying to be funny or meme-like? \\
\begin{itemize*}
	\item Yes
	\item No
\end{itemize*} \\

\noindent Was the video sponsored by a company? \\
\begin{itemize*}
	\item Yes
	\item No
\end{itemize*} \\

\noindent What types of information were contained inside the video's description?
\begin{itemize}[noitemsep,nosep]
	\item Additional information about items discussed in the video 
	\item Links to additional sources of information or citations for the video's contents
	\item Links to products
	\item Other \hrulefill
\end{itemize}
\vspace{\baselineskip}

\noindent Is the video aimed at parents or children? \\
\begin{itemize*}
	\item Parents
	\item Children
\end{itemize*} \\

\subsection{Parent-Oriented Content}
\textit{This section is only used if the video was aimed at parents.}\\

\noindent What was the topic of the video? (If it was just mentioned in passing, don't list it here)
\begin{itemize}[noitemsep,nosep]
	\item Setting up parental controls built into the device's OS 
	\item Installing a DNS-based content filter 
	\item Installing a content filter (not DNS-based) 
	\item Installing monitoring software 
	\item Preventing circumvention of content filtering or device monitoring
	\item Educating about general online safety concerns
	\item Educating about the ethics on filtering content and monitoring children
	\item Other \hrulefill
\end{itemize}
\vspace{\baselineskip}

\noindent Which types of devices were discussed? (If it was just mentioned in passing, don't list it here)
\begin{itemize}[noitemsep,nosep]
	\item Windows 
	\item macOS 
	\item Chromebooks 
	\item iOS 
	\item Android 
	\item Mobile devices or tablets (not specific to Android or iOS) 
	\item Gaming consoles 
	\item Other  \hrulefill
\end{itemize}
\vspace{\baselineskip}

\noindent \textit{Only displayed if preventing circumvention was one of the topics covered in the video}\\
For circumvention prevention, what strategies were discussed? (If it was just mentioned in passing, don't list it here)
\begin{itemize}[noitemsep,nosep]
	\item Locking down administrator rights on the child's devices 
	\item Restricting access to the router's admin functionality 
	\item Preventing VPN usage
	\item Preventing targeting avoidance (e.g., changing MAC address) 
	\item Other \hrulefill
\end{itemize}
\vspace{\baselineskip}

\noindent\textit{Only displayed if educating about ethics was one of the topics covered in the video}\\
\noindent What stance did the video take in regards to parents' right to protect their children and children's right for digital freedom? \\
\begin{itemize}[noitemsep,nosep]
	\item Pro parental rights 
	\item It depends / nuanced view / somewhere in the middle 
	\item Pro child rights 
\end{itemize}
\vspace{\baselineskip}

\noindent If there was a reason given for parents to need content filtering or device monitoring, did it involve any of the following?
\begin{itemize}[noitemsep,nosep]
	\item Social media can be dangerous and access to it needs to be limited 
	\item The internet can be distracting, and access to it needs to be limited (specific hours or total hours) 
	\item The internet is full of inappropriate material (e.g., pornography, cheating) 
	\item The Internet is full of malicious software that children accidentally download 
	\item There are malicious individuals online with whom children should not be allowed to communicate 
	\item Children are rebellious / bad / criminal and need to be controlled 
	\item Other  \hrulefill
\end{itemize}
\vspace{\baselineskip}

\subsection{Child-Oriented Content}
\textit{This section is only used if the video was aimed at children.}\\

\noindent What was the topic of the video? (If it was just mentioned in passing, don't list it here)
\begin{itemize}[noitemsep,nosep]
	\item Circumventing parental controls built into the device's OS 
	\item Circumventing an installed DNS-based content filter 
	\item Circumventing an installed content filter (not DNS-based) 
	\item Circumventing installed monitoring software 
	\item Educating about general online safety concerns (not parent-vs-child focused) 
	\item Educating about the ethics on filtering content and monitoring children 
	\item Other  \hrulefill
\end{itemize}
\vspace{\baselineskip}

\noindent Which types of devices were discussed? (If it was just mentioned in passing, don't list it here)
\begin{itemize}[noitemsep,nosep]
	\item Windows 
	\item macOS 
	\item Chromebooks
	\item iOS 
	\item Android 
	\item Mobile devices or tablets (not specific to Android or iOS) 
	\item Gaming consoles 
	\item Other  \hrulefill
\end{itemize}
\vspace{\baselineskip}

\noindent \textit{Only displayed if circumvention was one of the topics covered in the video}\\
For circumvention prevention, what strategies were discussed? (If it was just mentioned in passing, don't list it here)
\begin{itemize}[noitemsep,nosep]
	\item Changing settings on the child's device 
	\item Gaining access to the router's admin interface 
	\item Using a VPN 
	\item Employing target avoidance (e.g., changing the device's MAC address) 
	\item Other  \hrulefill
\end{itemize}
\vspace{\baselineskip}

\noindent\textit{Only displayed if educating about ethics was one of the topics covered in the video}\\
What stance did the video take in regards to parents' right to protect their children and children's right for digital freedom?
\begin{itemize}[noitemsep,nosep]
	\item Pro parental rights 
	\item It depends / nuanced view / somewhere in the middle 
	\item Pro child rights 
\end{itemize}
\vspace{\baselineskip}

\noindent If there was a reason given for needing to circumvent filtering or device monitoring, did it involve any of the following?
\begin{itemize}[noitemsep,nosep]
	\item Gaining increased access to social media 
	\item Getting around time-based restrictions on internet usage (specific hours or total hours) 
	\item Accessing whatever webpages the child wants 
	\item Communicating online with whoever the child wants 
	\item Abusive parents 
	\item Other  \hrulefill
\end{itemize}
\vspace{\baselineskip}

\subsection{Video Quality}

\noindent Please rate the quality of the video along the following three axes:\\
\begin{tabular}{l|ccc}
	& Yes & Somewhat & No \\
	\midrule
	Accurate & $\circ$ & $\circ$ & $\circ$ \\
	Comprehensive & $\circ$ & $\circ$ & $\circ$ \\
	Actionable & $\circ$ & $\circ$ & $\circ$ \\
\end{tabular}
\vspace{\baselineskip}

\noindent Please specifically identify what problems there were with how accurate the video was
\hrulefill \\

\noindent Please specifically identify what problems there were with how comprehensive the video was
\hrulefill \\

\noindent Please specifically identify what problems there were with how actionable the video was
\hrulefill \\

\noindent Did the video's title accurately describe the video's contents?
\begin{itemize*}
	\item Yes 
	\item No 
\end{itemize*} \\

\subsection{Final Notes}

\noindent Were there any other notes you would like to make about this video? \hrulefill